\def\alwaysmath#1{\ifmmode {#1} 
                  \else {$#1\mkern-5mu$} \fi}
\def\percc{\alwaysmath{\cm^{-3}}}
\def\kms{\alwaysmath{\km\sec^{-1}}}

\def\cmtwo{\alwaysmath{\,{\rm cm}^{-2}}}

\def\cm{\alwaysmath{\,{\rm cm}}}

\def\km{\alwaysmath{\,{\rm km}}}

\def\pers{\alwaysmath{\,{\rm \sec^{-1}}}}
\def\sec{\alwaysmath{\,{\rm s}}}

\def\mev{\alwaysmath{\,{\rm MeV}}}

\def\deg{\alwaysmath{\,^\circ}}

\def\pdeg{\alwaysmath{\,.\!\!^\circ}}

\def\gsim{\raisebox{-.7ex}{$\stackrel{\textstyle >}{\sim}$}}
\def\lsim{\raisebox{-.7ex}{$\stackrel{\textstyle <}{\sim}$}}

\def\ee#1{\alwaysmath{\times 10^{#1}}}

\def\lsun{\alwaysmath{\,L_\odot} }
\def\lfir{\alwaysmath{\,L_{FIR}}}

\def\nhh{\alwaysmath{n_{{}_{\rm H_2}}}}

\def\gray{\alwaysmath{\gamma {\rm -ray}}}
\def\grays{\alwaysmath{\gamma {\rm -rays}}}

\def\hi{H~{\small I}}

\def\htwo{\alwaysmath{{\rm H}_2}}

\def\C#1 {\alwaysmath{{}^{#1}{\rm C}}}

\documentstyle[12pt,aaspp4]{article}  

\begin{document}

\def\phib   {\hbox{$\mathaccent 22 \varphi$}}
\def\simlt  {\lower.5ex\hbox{$\; \buildrel < \over \sim \;$}}
\def\simgt  {\lower.5ex\hbox{$\; \buildrel > \over \sim \;$}}
\def\egr{\mbox{EGRET}\ }
\def\egrt{\mbox{EGRET\/}}
\def\degr{\hbox{$^\circ$}}
\def\ncr{\alwaysmath{n_{_{\!CR}}}}
\def\lsun{\alwaysmath{\,L_\odot}}

\title{Diffuse Gamma-Ray Emission from Starburst Galaxies and M31}

\author{Jan J. Blom, Timothy A. D. Paglione$^1$ and Alberto Carrami\~nana}
\affil{Instituto Nacional de Astrof\'{\i}sica, Optica y
Electr\'onica, Apartado Postal 216 y 51, 72000 Puebla, Pue., M\'exico;
blom@inaoep.mx}
\altaffiltext{1}{also Five College Radio Astronomy Observatory and Department
of Physics and Astronomy, Lederle Research Tower, University of Massachusetts,
Amherst, MA 01003}

\begin{abstract}

\noindent 

We present a search for high energy gamma-ray emission from 9 nearby
starburst galaxies and M31 with the Energetic Gamma-Ray Experiment
Telescope (EGRET) aboard the Compton Gamma-Ray Observatory (CGRO).
Though the diffuse gamma-ray emission from starburst galaxies was
suspected to be detectable, we find no emission from NGC 253, M82 nor
from the average of all 9 galaxies.  The 2$\sigma$ upper limit for
the EGRET flux above 100 MeV for the averaged survey observations is
1.8\ee{-8} ph\cmtwo\pers.  From a model of the expected radio and
\gray\ emission, we find that the magnetic field in the nuclei of
these galaxies is $>25 \mu$G, and the ratio of proton and electron
densities is $<400$.  The EGRET limits indicate that the rate of
massive star formation in the survey galaxies is only about an order
of magnitude higher than in the Milky Way.  The upper limit to the
\gray\ flux above 100 MeV for M31 is 1.6\ee{-8} ph\cmtwo\pers.  At the
distance of M31, the Milky Way flux would be over twice this value,
indicating higher \gray\ emissivities in our Galaxy.  Therefore, since
the supernova rate of the Milky Way is higher than in M31, our null
detection of M31 supports the theory of the supernova origin of cosmic
rays in galaxies.

\end{abstract}


\section{Introduction}

To date, the Large Magellanic Cloud (LMC) is the only normal galaxy
other than the Milky Way that has been seen in high energy, diffuse
\gray\ emission (Sreekumar et~al.\ 1992).  Most normal galaxies are
simply too distant, and their cosmic ray (CR) densities are too small
to have sufficient \gray\ emissivities.  Starburst galaxies, however,
have many supernovae due to their unusually high rates of massive star
formation. The supernovae produce copious CR electrons seen as bright,
extended synchrotron regions (e.g., Seaquist \& Odegard 1991; Carilli
et~al.\ 1992).  In addition, starburst nuclei typically contain high
masses of dense molecular gas (Young et~al.\ 1989; Solomon, Downes \&
Radford 1992; Helfer \& Blitz 1993; Paglione et~al.\ 1997).  Since CRs
are the likely cause of the high temperatures derived deep in the
cores of these clouds (Suchkov, Allen \& Heckman 1993), \gray\
production, through interactions between gas and CRs, should be
relatively high in starburst galaxies.  The challenge has been to
accumulate enough exposure time to detect the emission.

As the archetypal starburst galaxies (Rieke et~al.\ 1980), the nearby
galaxies NGC 253 and M82 have been the primary targets of \gray\
searches.  Both have high supernova rates of $\sim\,0.1$ yr$^{-1}$
derived from high resolution radio continuum images (van Buren \&
Greenhouse 1994; Ulvestad \& Antonucci 1997).  CO maps indicate
molecular gas column densities of $\sim 10^{23}$\cmtwo\ within their
inner kiloparsec (Paglione et~al.\ 1999; Carlstrom 1989), roughly 10
times that of the Milky Way (e.g. Sanders, Solomon, \& Scoville
1984).  Studies of their dense molecular gas indicate high average
densities of $\nhh<3\ee4$\percc\ in M82 to $10^6$\percc\ in NGC 253
(Paglione et~al.\ 1997).  Therefore, NGC~253 and M82 are expected to be
the brightest starburst galaxies in \grays.  However, previous EGRET
studies did not detect these objects at high energies (Sreekumar
et~al.\ 1994; Paglione et~al.\ 1996).

More CGRO data have become available in the meantime which should
improve sensitivity.  By some accounts (Pohl 1994), even these
starburst galaxies may not be luminous enough in \grays\ to be
detected individually by EGRET with the exposure currently available.
However, a combination of all observations obtained for a {\it
collection\/} of starburst galaxies may add up to a detectable
signal.  In this paper we present such a search from a survey of 9
nearby starburst galaxies. In addition we analyze \gray\ observations
of the nearby normal galaxy M31.

\section{Instrument and Data Analysis}

The \egr spark-chamber on board CGRO operates in the $30$ MeV to $30$
GeV energy range. The instrument is capable of producing circular
$\gamma$-ray images with typical radii of 30\degr.  The location
accuracy for point sources is about 0\pdeg5.  A detailed description
of the instrument and its performance is given by Hughes et al.\
(1980) and Kanbach et al.\ (1988).  The calibration properties are
described by Thompson et al.\ (1993).

Sky maps are obtained by applying a maximum-likelihood method (Mattox
et al.\ 1996), which provides flux estimates and statistical
significances of sources identified in the data. The instrumental
background of \egr is small compared to the measured celestial emission.
The celestial background in the \egr data is dominated by an isotropic
component, which is most likely diffuse extragalactic emission (further
discussed in Sreekumar et al.\ 1998). A flat background model which allows
for this isotropic emission is created for each observation and energy
range. For observations near the Galactic plane, we additionally take
into account Galactic diffuse emission by creating models for the
H{\small I} and CO distributions, and for the inverse-Compton (IC)
radiation.  These models are made by convolving maps of the gas and photon
distributions with an \egr point-spread function for each energy interval.
We apply H{\small I} maps based on various 21-cm surveys (combined into an
all-sky map; see Bloemen et~al.\ 1986), and the CO map of
Dame et al.\ (1987).  The IC distribution is derived from a cosmic ray propagation model by Strong \& Youssefi (1995).

We apply the maximum-likelihood analysis software originally
developed to analyze data from the COMPTEL $\gamma$-ray telescope
aboard CGRO.  This software has been augmented to analyze archival
\egr data (summarized in Blom 1997), and has been verified to give
flux values and significances consistent with published results of the
\egr team. 

We use individual archival viewing periods rather than the \egr
all-sky mosaic supplied by the CGRO Science Support Center, because
they cover a broader time span (to make our search as sensitive as
possible we need the maximum exposure available). The fluxes we derive
for several blazar sources near the Galactic plane are consistent
with the published \egr results, which indicates that our diffuse models
do not introduce significant deviations. For example, the blazar
2EG~J1730-130 at $(l,b) = (12\degr.03, 10\degr.81)$ was marginally detected
by EGRET above 100~MeV during a two-week pointing in February 1992 at a
flux level of $(4.2\pm 1.5)$\ee{-7} ph \cmtwo\pers\ (Thompson et~al.\ 1995),
or $(4.4\pm 1.5)$\ee{-7} ph \cmtwo\pers\ based on an improved analysis
(Mukherjee et~al.\ 1997). We find $(4.8\pm 1.5)$\ee{-7} ph \cmtwo\pers,
which is in good agreement with the two flux values published by the
EGRET team.

In order to combine different observations for a particular source, we
select data from all pointings taken within 30\degr\ of the source location
and coadd the count distributions and exposure maps separately. The coadding
procedure is straightforward only when the viewing periods have matching
coordinate systems. The EGRET archives usually provide data in Galactic
coordinates when observations are performed near the Galactic plane, otherwise viewing periods are given in equatorial coordinates. For most starburst
sources studied in this paper, observations with consistent coordinates
are available in the archives. However, in a few cases we are forced to
combine data in Galactic coordinates with data in equatorial coordinates.
Since we are only interested in the signal at one pixel (the source location),
we basically ignore the coordinate system. In other words, we shift the
origins of the data sets to the source location and subsequently coadd the
files assuming flat coordinates. The same procedure is used to coadd the
signals from a {\it collection\/} of sources.

Note that we have verified that the mosaics obtained from both coadding
procedures are reliable in the sense that analyses of combined blazar
observations yield flux values that agree with published results. When
we fit a point-source model to combined data at sky locations one or two
pixels away from the true source location (0\degr.5 to 1\degr\ off), we
obtain flux values that are lower than, but consistent with
the optimal value. This result shows that our coadding procedure is not
sensitive to small coordinate/binning mismatches between viewing periods.

\section{Observations and Results}

The observation program of CGRO consists of four phases.  Phases~I
to~III each consist of $\sim 1.5$ yr of observations.  Phase~IV
is subdivided into annual cycles, starting with Cycle IV.  Data from
Phase I up to Phase IV/Cycle V, covering the period 1991--1996, are
used in this work.  Table~\ref{tab.sou} shows the relevant EGRET
observations for each galaxy in our survey. Specific
observing dates of the CGRO phases and viewing periods are given
elsewhere (e.g., Gehrels et~al.\ 1994).

The survey consists of ten starburst galaxies selected by distance
($\lsim$ 10 Mpc), IRAS far infrared (FIR) luminosity 
($>10^{9}$\lsun, Young et~al.\ 1989), and angular distance from the
Galactic plane ($b>10\deg$).  We omit one candidate, NGC~891, because
of its proximity to the EGRET source 2EG~J0220+4228 (Thompson et~al.\ 1995;
Mattox et~al.\ 1997). The large spiral galaxy M31 is analyzed separately
as well.  Though its FIR luminosity (and therefore the supernova rate) is
an order of magnitude lower than a starburst galaxy's, M31 may be close
enough to make up for its lower \gray\ emissivity.

\subsection{NGC 253, M82 and M31}

All observations listed in Table~\ref{tab.sou} for NGC~253 and M82
are combined separately.  For each source, maximum-likelihood ratio
maps are created in six standard EGRET energy ranges (70--100 MeV,
100--150 MeV, 150--300 MeV, 300--500 MeV, 0.5--1 GeV, and 1--2 GeV),
and in one broad energy range covering all photon energies $>
100$ MeV.  Flux values are derived by simultaneously fitting
point-source models at the locations of the starburst galaxy and all
respective neighboring EGRET catalog sources within $25\degr$ on the
sky (Thompson et~al.\ 1995, 1996).

None of the likelihood maps show evidence for emission from NGC~253 or
M82.  Upper limits ($2\sigma$) obtained from model fitting are shown
in Figure~\ref{fig.gammas}.  Note that no upper limit is derived
for M82 at 1--2 GeV due to limited statistics.  For NGC~253 we find
a $2\sigma$ upper limit to the \gray\ emission above 100 MeV of $3.4\times
10^{-8}$ ph cm$^{-2}$ s$^{-1}$.  For M82 this limit is $4.4\times 10^{-8}$
ph cm$^{-2}$ s$^{-1}$.  Note that both limits are well below previous
estimates (Sreekumar et~al.\ 1994; Paglione et~al.\ 1996), which
further constrains emission models (\S\ref{model}).

We apply the same analysis technique for M31 at $> 100$ MeV only.
No evidence for M31 is found in this energy range. We find
a 2$\sigma$ upper limit for the flux above 100 MeV of 
1.6\ee{-8} ph \cmtwo\pers. This value is below the estimate of
\"Ozel \& Berkhuijsen (1987).

\subsection{Combining All Available Starburst Galaxy Observations}

In order to create a ``supermosaic'' for all starburst galaxies
listed in Table~\ref{tab.sou}, the observations for each source are
first combined separately.  The resulting submosaics are then
coadded such that the locations of all starburst galaxies match in
one pixel (with arbitrary coordinates). In addition, the event
distributions of the submosaics are scaled to ensure that each
starburst galaxy contributes equally in exposure to the supermosaic.
All background models are treated similarly.

We create maximum-likelihood ratio images for the supermosaic in six
standard EGRET energy ranges between 70 MeV and 2 GeV, and for all
photon energies $> 100$ MeV.  These images show significant excesses
that can be identified with the source signals of strong EGRET blazars
which remain visible in the supermosaic even after adding noise from
observations that do not contain these sources.  However, no evidence
for \gray\ emission at the shared location of the starburst galaxies
is found.  Upper limits derived from simultaneous model fitting taking
the EGRET blazars into account are shown in the right panel of
Figure~\ref{fig.gammas}.  The $2\sigma$ upper limit we derive for the
{\it average\/} \gray\ emission above 100 MeV from starburst galaxies
is $1.8\times 10^{-8}$ ph cm$^{-2}$ s$^{-1}$.

\section{The Expected Gamma-Ray Emission from Starburst Galaxies}
\label{model}

To calculate the \gray\ flux from a starburst galaxy, we model the
CR production and radiation mechanisms for NGC 253 and M82.  A full
description of the model is given in Paglione et~al.\ (1996).  In
short, we assume a power law form for the initial CR injection
spectrum $Q=KE^{-s}$, and allow it to evolve to a steady state. The
ambient density (with a main contribution from \htwo\ in galactic nuclei),
magnetic field and photon field determine the energy loss rates due to
bremsstrahlung, synchrotron and IC radiation, as well as secondary
particle production.  The model yields steady state primary and
secondary electron, and proton energy distributions, which are used to
calculate synchrotron and \gray\ spectra.  The normalization of the
proton injection spectrum is estimated by comparing the modeled and
observed synchrotron radio spectra.  This normalization corresponds to
the total power per unit time transferred by supernovae into cosmic rays
within a given volume,
\begin{equation}
\int^\infty_{E_{min}} Q E\ dE = \eta P \Psi/V\ \ .
\end{equation}

\noindent
Here $\eta$ is the efficiency of energy transfer from a supernova to
CRs, $P \approx 10^{51}$ ergs is the supernova energy, $\Psi$ is the
supernova rate, and $V$ is the volume.  Solving for the
normalization $K$, we find (for a minimum energy $E_{min}$ and a power
law slope of 2.2),
\begin{equation}
K = \frac{\eta P \Psi}{V}\, 0.2\ E_{min}^{0.2}\ \ .
\label{eq.k}\end{equation}

\noindent
A minimum CR energy, $E_{min}\,\gsim\, 2 m_p v_s^2$ (Bell 1978), is
required for acceleration in a shock front. It is roughly a few MeV
for a shock velocity of $10^4$\kms.  To generate \gray\ fluxes, the
model results are integrated over volume and divided by 4$\pi D^2$,
where $D$ is the galaxy distance.

\subsection{NGC 253 and M82}

For NGC 253, we choose a disk-shaped starburst region 70 pc thick,
with a radius of 325 pc (Paglione et al.\ 1999), and $\Psi = 0.08$
yr$^{-1}$ (Ulvestad \& Antonucci 1997).  For M82, $\Psi = 0.1$
yr$^{-1}$ (van Buren \& Greenhouse 1994), and the disk is 50 pc
thick with a radius of 355 pc (Shen \& Lo 1995).  Paglione
et~al.\ (1996) tested for \gray\ emission outside the starburst region
and found it to be negligible, so we ignore any interactions outside the
nucleus.  The calculated synchrotron spectra resulting from the
model CR electron distributions match the shape of the observed radio
spectra from the nuclei of NGC 253 and M82 (Carilli 1996; Carlstrom \&
Kronberg 1991) given a photon field of 200 eV\percc\ and a diffusion
time scale of 10 Myr (cf.\ Paglione et~al.\ 1996).  The average gas
densities for NGC 253 and M82 are 300 and 100 \percc, respectively.
Emission measures of $\sim 3\ee{5}$ and 10$^6$ cm$^{-6}$ pc are
required to match the free-free component and thermal absorption (low
frequency turn-over) of the spectra from NGC 253 and M82, respectively
(cf., Carilli 1996; Carlstrom \& Kronberg 1991).

On the whole, the density and photon field are well constrained by the
radio spectra.  The density range is limited by the slope of the radio
spectrum.  Higher densities increase the CR losses due to collisions
with matter, thus reducing the number of low energy CRs, resulting in
a flatter synchrotron spectrum.  (Note that thermal absorption only
affects the spectrum at low frequencies and is easily distinguished
from the inherent slope of the underlying emission.)  The
volume-averaged densities that best match the radio spectra agree well
with those found from molecular line studies (Paglione et~al.\ 1997).
The photon field is determined by the FIR luminosity and, to a much
lesser degree, by the synchrotron slope.  A steeper slope indicates a
higher photon field since more high energy electrons are lost due to
IC radiation.

The absolute level of the model synchrotron spectrum is determined by
the magnetic field $B$, the ratio of proton and electron densities
$N_p/N_e$, and $\eta$.  Though the magnetic field, derived from
radio data, is most likely $\gsim\, 50 \mu$G in NGC 253 and M82 (Carilli
1996; Seaquist \& Odegard 1991), $N_p/N_e$ and $\eta$ are unknown.
The $N_p/N_e$ ratio can be anywhere between 20 and 2000, though values
of 50 to a few hundred appear most likely for the Milky Way (Bell 1978).
Values for the efficiency of energy transfer between 10 and 50\%\ have
been found from models of the evolution of supernova blast waves
(Markiewicz, Drury \& V\"olk 1990). Fortunately, the \gray\ flux depends
only very weakly on $B$ and $N_p/N_e$.  Therefore, $\eta$ is the most
important parameter needed to predict the \gray\ flux, and it is determined
from the radio spectrum normalization.

Five models (A--E, Table~\ref{tab.models}) are shown with the EGRET
upper limits in Figure~\ref{fig.gammas}.  The \gray\ spectra are
summations of IC, bremsstrahlung and neutral pion decay radiation.
Near 100 MeV, the emission is dominated by neutral pion decays as seen
by the so-called ``pion bump.''  Though all the models lie below the
M82 limits, model~A predicts a \gray\ flux above 100 MeV for NGC 253
higher than measured.

\subsection{Starburst Galaxy Survey}\label{sbg}

From their integrated radio spectra and molecular gas properties, the
other galaxies in the survey are relatively similar to NGC 253 and
M82.  We assume that the major difference in the normalization
between starburst galaxies is the supernova rate and distance
(Equation~\ref{eq.k}).  We estimate the supernova rate from the FIR
luminosity (van Buren \& Greenhouse 1994).  For NGC 4945, a starburst
with very high gas densities (Jackson et~al.\ 1995), we use the model
fit to NGC 253.  For the other starburst galaxies, we use the model
for M82.  The expected average \gray\ spectra from models A--E are shown
in Figure~\ref{fig.gammas} with the EGRET upper limits.  Unlike the
models for NGC 253, which are constrained by the EGRET points near the
pion bump, these results are limited by the data near 1 GeV.

\section{Discussion}

Though many of the model fluxes lie below the EGRET limits, all but
one or two models may be eliminated. The low magnetic field in model~A
contradicts minimum energy estimates from radio data (Beck et~al.\ 1994;
Rieke et~al.\ 1980). Models~A and B both have very high transfer
efficiencies, and their \gray\ spectra are above or very near
near the 2$\sigma$ limits in a few energy bands for NGC 253 and the
starburst galaxies. With model~B eliminated, we conclude that $N_p/N_e < 
400$, as in the Milky Way. Note that model~B may still be valid for 
M82, though we have no reason to believe that the CR population would
be so different in this particular galaxy.

The magnetic field used for model~C agrees with previous
calculations, and the $N_p/N_e$ ratio of 100 matches that observed for
the Milky Way near 1 GeV.  The integrated fluxes and spectra all fall
below the EGRET limits.

After reviewing the calculation of the minimum energy magnetic field,
we find model~D to be a poor solution.  From the equation for magnetic
field strength in Beck et al.\ (1994), we find a gas volume filling
factor in NGC 253 of $\sim 5\ee{-3}$ for model~D.  This result implies
ionized cloud densities of roughly $10^{4.5}$\percc.  Since these
densities approach those found for the dense molecular medium in NGC
253 (Paglione et~al.\ 1999), model~D may not be appropriate due to its
low volume filling factor.  Assuming similar magnetic fields in M82
and the starburst galaxies, model~D is most likely not valid for them
either.

Another argument against model~D comes from comparing its predictions
with the expected Milky Way flux.  The \gray\ luminosity of the Milky Way
measured by EGRET above 100 MeV is roughly 2\ee{42} ph\pers
(S.~Hunter, private communication), and its FIR luminosity is
7\ee{9}\lsun (Cox \& Mezger 1989). When we scale the Milky Way \gray\
luminosity to the average starburst FIR luminosity (which is roughly
4 times higher) and distance (7.4 Mpc), we would observe a \gray\ flux
of $F_\gamma(E>100\mev) \sim 1.3\ee{-9}$ ph \cmtwo\pers. This is nearly
the flux predicted for the survey sources with model~D. Since the \gray\
emissivity should be higher in starburst galaxies due to their larger
average gas densities, this simple scaling most likely underestimates
the true flux.  We therefore consider model~D at best a lower limit
prediction.

Model E is eliminated since the predicted fluxes all fall below the
scaled estimate from the Milky Way.  Further, the \gray\ flux of the
LMC (Sreekumar et~al.\ 1992), when similarly scaled, is consistent
with, or slightly above, the fluxes from model~E.  The \gray\
emissivity of the LMC is much lower than in starburst galaxies. At a
distance of 50 kpc, even the \gray\ flux of the Milky Way is already 
35 times higher than in the LMC. Therefore, the fluxes predicted by
model~E are clearly too low.  Note that this result does not strictly
eliminate $N_p/N_e\le 50$ -- with this ratio we can obtain fluxes that
are similar to model~D given $\eta=0.13$ and $B=50 \mu$G.

Based on our new analysis and EGRET limits, we expect the flux
estimates from model~C to be the most realistic (with model~D as a
lower limit).  These models imply $N_p/N_e < 400$, similar to the
Galactic value for CR energies $\gsim\, 1$ GeV.  The energy from
supernovae is also efficiently transferred to CRs ($\eta \sim
20$--50\%).  The survey sources have magnetic field strengths
$>25\,\mu$G.

The EGRET limits also constrain the relative supernova (or massive
star formation) rate in the starburst galaxies.  At the average
distance of our survey sources, the flux above 100 MeV from the Milky
Way would be 3\ee{-10} ph \cmtwo\pers.  Compared with the observed
starburst upper limit, we find $\Psi_{SBG}/\Psi_{MW} < 60$.  This
limit is conservative since not even a hint of starburst emission
is found in the EGRET maps.  The integrated fluxes from models~C and D
indicate $\Psi_{SBG}/\Psi_{MW}\,\lsim\, 20$ and the difference in FIR
luminosities implies $\Psi_{SBG}/\Psi_{MW}\,\gsim\, 4$.  Given the
uncertainties in $\eta$, we estimate that the star formation rate in
starburst galaxies is only about an order of magnitude higher than in
the Milky Way.  Studies at other wavelengths also indicate star
formation rates elevated by this magnitude (e.g., Kennicutt 1998).

\subsection{M31 and the Supernova Origin of Cosmic Rays}

It is interesting to note that at the distance of M31, the flux of
the Milky Way above 100 MeV would be 4\ee{-8} ph \cmtwo\pers.  This
flux is over twice the EGRET limit for M31.  Therefore, the Galactic
\gray\ emissivity must be higher than in M31.  Three main differences
between our Galaxy and M31 can explain this result.  First, the
supernova rate in the Milky Way is roughly 7 times higher than in M31,
as can be estimated from the FIR luminosity. This difference is probably
even more pronounced because much of the FIR emission in M31 originates
from Population II stars in the nucleus rather than from massive star
forming regions (Dame et~al.\ 1993).  Therefore, if CRs are in fact
accelerated in the shock fronts of supernovae, the CR density in the
Milky Way, and thus its \gray\ emissivity, should be higher.  Second,
the mass of \hi\ and \htwo\ in the Milky Way is at least 6 times that
of M31 (e.g., Dame et~al.\ 1993; Clemens, Sanders \& Scoville 1988).
Third, unlike the starburst galaxies and the Milky Way, M31 lacks a
massive concentration of dense molecular gas in its nucleus (though
some unexplained radio emission is seen there). This implies a lower
secondary particle production in M31 (including neutral pions) which
means a lower diffuse \gray\ flux.  However, the \gray\ emissivity
does not depend very strongly on density, so the difference in supernova
rates is the most important factor determining the high energy emission
from galaxies.  We conclude that our null detection of hard \gray\ emission
from M31, compared with the predicted Milky Way flux at the same distance,
supports the theory of the supernova origin of CRs.

\subsection{Possible Point-source Contributions}

Throughout the paper we have neglected point-source contributions
to the high-energy emission from starburst galaxies, M31, and the Milky
Way. In principle, a large population of relatively weak \gray\ sources
may add significantly to the total high-energy emission of a galaxy. 
Studies on the nature of the many unidentified EGRET sources suggest that
supernova remnants (without an embedded pulsar), X-ray binaries and flare
stars are all possible, but not yet well-established sources of high-energy \grays\ (Mukherjee, Grenier \& Thompson 1997). Pulsars are the only firmly identified small-scale objects that are known to emit \grays\ up to
GeV energies. Therefore, we calculate the \gray\ emission from a
Galactic ensemble of pulsars to estimate the possible contribution of
such a source population to the overall \gray\ luminosity in a galaxy.

EGRET has detected 7 pulsars above $100$ MeV in the Galaxy which show an
increasing efficiency in converting rotational energy of the spinning
neutron star into \grays\ versus pulsar age. A second obvious trend
is an increasing spectral hardening in the \gray\ regime versus age
(Fichtel \& Trombka 1997). Their typical pulsed \gray\ spectra can be
described by power-laws which cut off or break at $1$--$10$ GeV
(Thompson et~al.\ 1997). All {\it radio\/} pulsars in the (fairly complete)
Princeton Pulsar catalog (Taylor, Manchester \& Lyne 1993) may generally
have similar \gray\ emission properties. When we assume that the radio and
\gray\ emission cones of pulsars have the same aperture, we can easily
estimate an integrated pulsar contribution to the high-energy emission of
the Galaxy above 100 MeV. We find a maximum contribution of $2\%$ when we
adopt a long-lived pulsar activity ($\sim 10^7$ yr) and an average spectral
cut-off at low \gray\ energies ($\sim 2$ GeV). This result indicates that
pulsars add an insignificant fraction to the total \gray\ luminosity of
the Galaxy above 100 MeV. This is in agreement with the findings of
Pohl et~al.\ (1997), who conclude that only for energies $> 1$ GeV we
may find a significant contribution (up to $18\%$) in selected sky regions.

The flux values listed in the EGRET catalogs imply that the integrated
\gray\ emission from unidentified sources is comparable to that from pulsars.
If all the EGRET unidentifieds are Galactic sources of some type other
than pulsars (which is in fact unlikely), then we have a maximum total
point-source contribution of $4\%$ to the overall emission $> 100$ MeV
in the Galaxy. Since we have no reason to expect a more dominant contribution
from pulsars (and other point sources) in starburst galaxies or M31, we
conclude that it is justified to neglect point-source contributions in our
model predictions.

\section{Conclusions}

Our combined EGRET observations are the most sensitive measurements
of starburst galaxies ever performed at hard \gray\ energies.
However, no MeV or GeV emission from starburst galaxies is
identified.  Nevertheless, we have shown that the derived EGRET upper
limits are useful for constraining models of CR and \gray\ production
in galaxies.  The modeling implies large average gas densities,
magnetic fields, and photon fields.  We find $50\,\lsim\, N_p/N_e <
400$, similar to the Milky Way value, and a high efficiency of energy
transfer from supernovae to CRs (20--50\%). These values agree well with
radio observations.

The \gray\ emissivity of M31 is at most half that of the Milky Way.
This result supports the supernova origin of CRs because of the lower
supernova rate of M31.

According to our model predictions, an instrument with an order of
magnitude better sensitivity than EGRET should be able to detect NGC
253 and M82 in a one-year all-sky survey.  The Gamma-ray Large Area
Space Telescope (GLAST; e.g., Michelson 1996), with a projected
point-source sensitivity in one year of 5\ee{-9} ph \cmtwo\pers, may
have this capability. The GLAST pair conversion telescope is being
designed to image the 10 MeV -- 300 GeV sky. Its launch is expected
in the first decade of the next century.

\acknowledgements

This research has made use of the NASA/IPAC Extragalactic Database
(NED) which is operated by the Jet Propulsion Laboratory, Caltech,
under contract with the National Aeronautics and Space Administration.
J. J. B. acknowledges computational support from the Space Research
Organization Netherlands (SRON) in Utrecht, The Netherlands. We thank
S. Bloom and S. Hunter for helpful discussions.

\clearpage

\figcaption[gammas.ps]
{The expected \gray\ flux from (left to right) NGC 253, M82 and the
starburst galaxies (SBGs), shown with the EGRET 2$\sigma$ upper limits
(dots).  Model parameters are listed in
Table~\protect\ref{tab.models}.
\label{fig.gammas}} 

\begin{deluxetable}{lccccl}
\tablewidth{0pt}
\tablecaption{Starburst Galaxies and EGRET Observations.
\label{tab.sou}}
\tablehead{
Galaxy & $D$ & $l$ & $b$ & \lfir & CGRO Viewing Periods \\
 & (Mpc) & &  & (10$^{10}$\lsun)&
}
\startdata
M31    & \phn0.67 & 121\pdeg1 & $-$21\pdeg6\phm{$-$} & \phm{$^a$}0.1$^a$ &  26.0, 28.0, 34.0, 37.0, 211.0, \nl
       &     &      &      &            & 325.0, 401.0, 425.0, 530.0 \nl
       &     & & & & {\bf Total Exposure:} $0.6\times 10^9$ cm$^2$ s \nl
IC 342 & 1.8 & 138.2 & 10.6 & 0.2 & 15.0, 18.0, 31.0, 211.0, 216.0, \nl
       &     &      &      &            & 319.0, 319.5, 325.0, 401.0, 411.0, \nl
       &     &      &      &            & 411.5, 518.5, 530.0               \nl
       &     & & & & {\bf Total Exposure:} $1.3\times 10^9$ cm$^2$ s  \nl
M82    & \phn3.25 & 141.4    & 40.6   & 3.2 & 18.0, 22.0, 216.0, 227.0, 228.0, 319.0,\nl
       &          &            &          &     & 319.5, 411.0, 411.5, 418.0, 518.5 \nl
       &          & & & & {\bf Total Exposure:} $1.8\times 10^9$ cm$^2$ s  \nl
NGC 253& 3.4 & \phn97.4 &$-$88.0\phm{$-$}& 2.8 & 9.0, 13.5, 404.0, 428.0      \nl
       &     & & & & {\bf Total Exposure:} $0.6\times 10^9$ cm$^2$ s  \nl
M83    & 5.4 & 314.6  & 32.0 & 2.0 & 12.0, 23.0, 32.0, 207.0, 208.0,  \nl
       &     &           &     & & 215.0, 217.0, 316.0, 405.0, 405.5, \nl
       &     &           &     & & 408.0, 424.0, 511.5   \nl
       &     & & & & {\bf Total Exposure:} $1.1\times 10^9$ cm$^2$ s  \nl
NGC 2903 & 7.0 & 208.7 & 44.5 & 0.8 & 40.0, 322.0, 326.0, 403.5    \nl
       &       &     & & & {\bf Total Exposure:} $0.6\times 10^9$ cm$^2$ s  \nl
NGC 4945 & 7.0 & 305.3 & 13.3 & \phm{$^a$}8.5$^a$& 12.0, 14.0, 23.0, 27.0, 32.0, \nl
       &       &          &     &  & 207.0, 208.0, 215.0, 217.0, 230.5,  \nl
       &       &          &     &  & 303.0, 314.0, 315.0, 316.0, 402.0,  \nl
       &       &          &     &  & 402.5, 424.0, 522.0, 531.0  \nl
       &       & & & & {\bf Total Exposure:} $1.4\times 10^9$ cm$^2$ s  \nl
M51    & 9.7 & 104.8 & 68.6 & 3.1 & 4.0, 22.0, 218.0, 222.0, 418.0, 515.0  \nl
       &     & & & & {\bf Total Exposure:} $0.6\times 10^9$ cm$^2$ s  \nl
NGC 6946 & 11.0\phn & \phn95.7  & 11.7 & 4.8 & 2.0, 34.0, 203.0, 212.0, 302.0,  \nl
       &            &             &     &  & 303.2, 303.7, 318.1, 401.0, 530.0 \nl
       &            & & & & {\bf Total Exposure:} $0.9\times 10^9$ cm$^2$ s  \nl
NGC 3628 & 11.5\phn & 169.4  & 13.9 & 1.9 & 3.0, 4.0, 11.0, 204.0, 205.0, 206.0, \nl
       &            &          &     &  & 222.0, 304.0, 305.0, 306.0, 307.0, \nl
       &            &          &     &  & 308.0, 308.6, 311.0, 311.6, 312.0, \nl
       &            &     &     &  & 313.0, 322.0, 326.0, 511.0, 515.0\nl
       &            & & & & {\bf Total Exposure:} $1.2\times 10^9$ cm$^2$ s  \nl

\tablenotetext{a}{Adapted from Rice et~al.\ (1988)}
\enddata
\end{deluxetable}

\begin{deluxetable}{lcccccc}
\tablewidth{0pt}
\tablecaption{Parameter values for starburst emission models.
\label{tab.models}}
\tablehead{
Model & $B$ & $N_p/N_e$ & $\eta$ & \multicolumn{3}{c}{$F(E>100 \mev)
\ 10^{-9}$ ph\cmtwo\pers} \\
\cline{5-7} \\
 & ($\mu$G) & & & NGC 253 & M82 & SBGs 
}
\startdata
A \dotfill & \phn25 & \phn50 & 0.90 & 36.5 & 37.5 & 14.4 \nl
B \dotfill & \phn50 & 400 & 0.73 & 24.1 & 27.6 & 10.0 \nl
C \dotfill & \phn50 & 100 & 0.43 & 15.6 & 16.7 & \phn6.3 \nl
D \dotfill & 100    & 100 & 0.15 & \phn5.3 & \phn6.1 & \phn2.2 \nl
E \dotfill & 100 & \phn50 & 0.05 & \phn2.0 & \phn2.2 & \phn0.8 \nl
\enddata
\end{deluxetable}

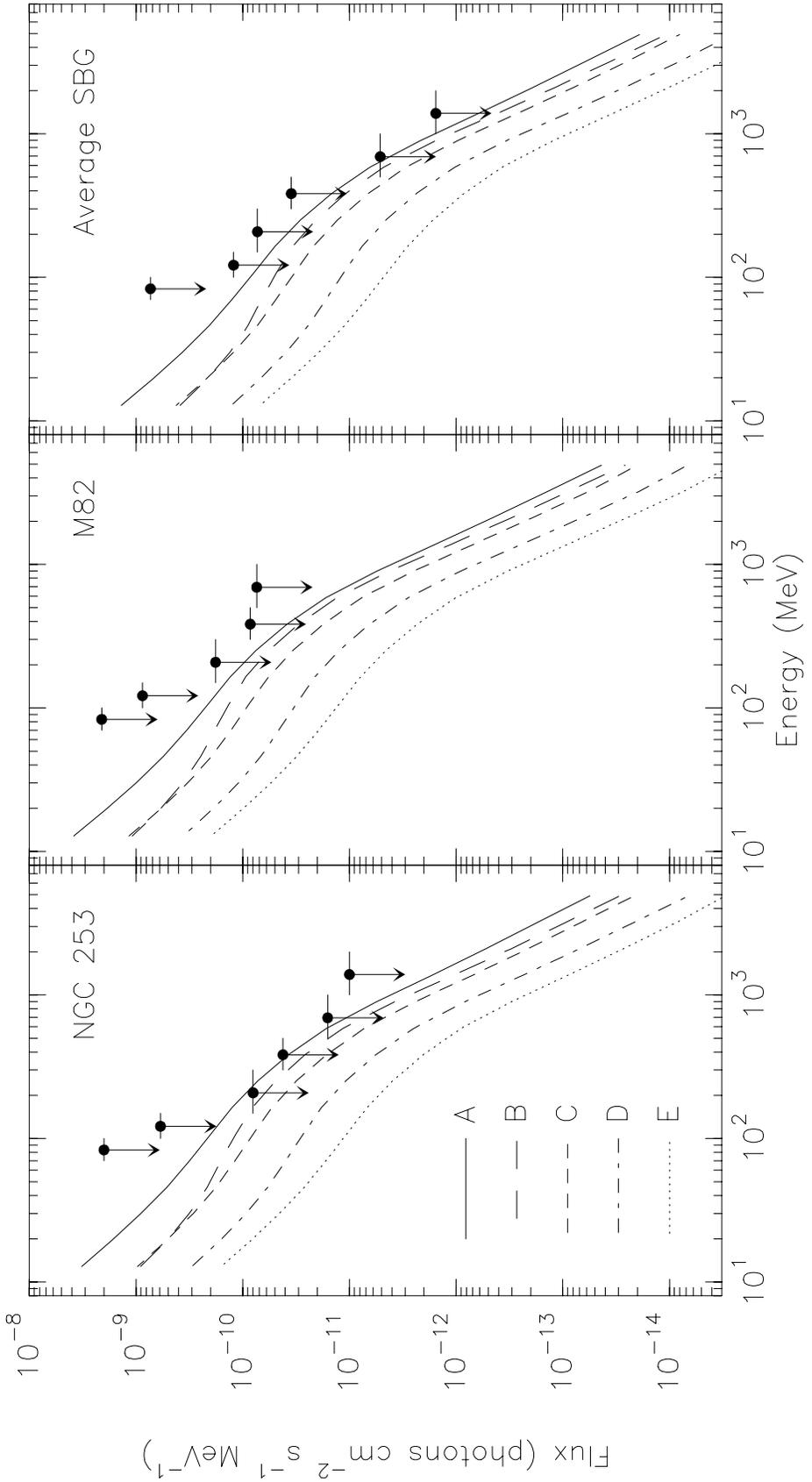
\end{document}